\documentclass[a4paper,11pt]{article}

\usepackage{jinstpub} 

\usepackage{color}

\usepackage{lineno}

\title{Development of large-volume $^{130}$TeO$_2$ bolometers for the CROSS $2\beta$ decay search experiment}

\author[a]{F.T.~Avignone~III,}
\author[b]{A.S.~Barabash,}
\author[c]{V.~Berest,}
\author[d]{L.~Berg{\'e},}
\author[e,f,g]{J.M.~Calvo-Mozota,}
\author[h,i]{P.~Carniti,}
\author[d]{M.~Chapellier,}
\author[j]{I.~Dafinei,}
\author[k,l]{F.A.~Danevich,}
\author[d]{L.~Dumoulin,}
\author[m]{F.~Ferella,}
\author[c]{F.~Ferri,}
\author[d]{A.~Gallas,}
\author[d]{A.~Giuliani,}
\author[h]{C.~Gotti,}
\author[c]{P.~Gras,}
\author[m]{A.~Ianni,}
\author[d]{L.~Imbert,}
\author[c]{H.~Khalife,}
\author[k]{V.V.~Kobychev,}
\author[b]{S.I.~Konovalov,}
\author[d]{P.~Loaiza,}
\author[d]{P.~de~Marcillac,}
\author[d]{S.~Marnieros,}
\author[d]{C.A.~Marrache-Kikuchi,}
\author[n,o]{M.~Martinez,}
\author[m]{S.~Nisi,}
\author[c]{C.~Nones,}
\author[d]{E.~Olivieri,}
\author[n]{A.~Ortiz de Sol\'orzano,}
\author[d]{Y.~Peinaud,}
\author[h]{G.~Pessina,}
\author[d]{D.V.~Poda,}
\author[d]{Ph.~Rosier,}
\author[d]{J.A.~Scarpaci,}
\author[k,m]{V.I.~Tretyak,}
\author[b]{V.I.~Umatov,}
\author[k]{M.M.~Zarytskyy,}
\author[c]{and A.~Zolotarova}

\affiliation[a]{Department of Physics and Astronomy, University of South Carolina, Columbia, 29208 South Carolina, USA}
\affiliation[b]{National Research Center Kurchatov Institute, Kurchatov Complex of Theoretical and Experimental Physics, 117218 Moscow, Russia}
\affiliation[c]{IRFU, CEA, Université Paris-Saclay, 91191 Saclay, France}
\affiliation[d]{Universit\'e Paris-Saclay, CNRS/IN2P3, IJCLab, 91405 Orsay, France}
\affiliation[e]{Laboratorio Subterr\'aneo de Canfranc, 22880 Canfranc-Estaci\'on, Spain}
\affiliation[f]{Escuela Superior de Ingenier\'ia, Ciencia y Tecnolog\'ia, Universidad Internacional de Valencia -- VIU, 46002 Valencia, Spain}
\affiliation[g]{Escuela Superior de Ingenier\'ia y Tecnolog\'ia, Universidad Internacional de La Rioja, 26006 Logro\~no, Spain}
\affiliation[h]{INFN Sezione di Milano-Bicocca, I-20126 Milan, Italy}
\affiliation[i]{Universit\`{a} di Milano-Bicocca, I-20126 Milan, Italy} 
\affiliation[j]{INFN Sezione di Roma, I-00185 Rome, Italy}
\affiliation[k]{Institute for Nuclear Research of NASU, 03028 Kyiv, Ukraine}
\affiliation[l]{INFN Sezione di Roma Tor Vergata, I-00133 Rome, Italy}
\affiliation[m]{INFN Laboratori Nazionali del Gran Sasso, I-67100 Assergi (AQ), Italy}
\affiliation[n]{Centro de Astropart\'iculas y F\'isica de Altas Energ\'ias, Universidad de Zaragoza, 50009 Zaragoza, Spain}
\affiliation[o]{ARAID Fundaci\'on Agencia Aragonesa para la Investigaci\'on y el Desarrollo, 50018  Zaragoza, Spain}

\emailAdd{andrea.giuliani@ijclab.in2p3.fr} 
\emailAdd{denys.poda@ijclab.in2p3.fr}

\abstract{  
We report on the development of thermal detectors based on large-size tellurium dioxide crystals (45 $\times$ 45 $\times$ 45 mm), containing tellurium enriched in $^{130}$Te to about 91\%, for the CROSS double-beta decay experiment. A powder used for the crystals growth was additionally purified by the directional solidification method, resulting in the reduction of the concentration of impurities by a factor 10, to a few ppm of the total concentration of residual elements (the main impurity is Fe). The purest part of the ingot (the first $\sim$200 mm, about 80\% of the total length of the cylindrical part of the ingot) was determined by scanning segregation profiles of impurities and used for the $^{130}$TeO$_2$ powder production with no evidence of re-contamination. The crystal growth was verified with precursors produced from powder with natural Te isotopic composition, and two small-size (20 $\times$ 20 $\times$ 10 mm) samples were tested at a sea-level laboratory showing high bolometric and spectrometric performance together with acceptable $^{210}$Po content (below 10 mBq/kg). This growth method was then applied for the production of six large cubic $^{130}$TeO$_2$ crystals and 4 of them were taken randomly to be characterized at the Canfranc underground laboratory, in the CROSS-dedicated low-background cryogenic facility. Two $^{130}$TeO$_2$ samples were coated with a thin, $O$(100~nm), metal film in form of Al layer (on 4 sides) or AlPd grid (on a single side) to investigate the possibility to tag surface events by pulse-shape discrimination. Similarly to the small natural precursors, large-volume $^{130}$TeO$_2$ bolometers show high performance and even better internal purity ($^{210}$Po activity $\sim$ 1 mBq/kg, while activities of $^{228}$Th and $^{226}$Ra are below 0.01~mBq/kg), satisfying requirements for the CROSS and, potentially, next-generation experiments.
}


\arxivnumber{2406.01444} 

\begin{document}
\maketitle
\flushbottom

\section{Introduction}
\label{sec:intro} 

The search and investigation of spontaneous nuclear transitions like double-beta ($2\beta$) processes, which are expected to occur rarely or even to be forbidden according to the Standard Model (SM) of particle physics, have represented a great interest for nuclear and particle physics until recently \cite{GomezCadenas:2023,Agostini:2023,Workman:2022,APPEC_DBD:2020}. 
The mostly searched $2\beta$ processes (driven by high transition energy and a relatively large presence in nature of the isotopes of interest \cite{Tretyak:2002}) are decays in which two electrons appear in the final state accompanied or not with the emission of two anti-neutrinos, corresponding to two-neutrino ($2\nu$) and neutrinoless ($0\nu$) $2\beta$ decays respectively. 
The former process has been detected for a dozen nuclei with half-lives of 10$^{18}$--10$^{24}$ yr \cite{Barabash:2020}, while the latter one remains unobserved and the most stringent half-life limits reach 10$^{24}$--10$^{26}$ yr level \cite{Agostini:2023}. The observation of $0\nu2\beta$ decay would help to scrutinize many open questions of particle physics such as the lepton-number conservation, the absolute value of neutrino mass and its origin (Majorana or Dirac particle), the admixture of right-handed currents in weak interactions, the existence of majorons and bosonic neutrinos, the imbalance between matter and anti-matter (via leptogenesis) and other effects beyond the SM \cite{Agostini:2023}. 

Among the candidates studied, $^{130}$Te stands out as one of the few extensively researched $2\beta$-active isotopes, thanks to: 
i) a relatively high transition energy ($Q_{2\beta}$ = 2527 keV \cite{Wang:2021a}); 
ii) a considerable presence in natural tellurium ($\delta$ = 34\% \cite{Meija:2016});
iii) the possibility to grow large-volume, high-quality, and radiopure tellurium dioxide (TeO$_2$) crystals, i.e. a dielectric material with a high tellurium content suitable for the use in low-temperature detectors of rare events \cite{Chu:2006,Barucci:2001,Alessandria:2012,Alduino:2018a}. 
Thanks to the above-listed features, TeO$_2$ is among a few materials with the longest history of applications to $2\beta$ searches. Indeed, initial attempts were made in the early 1990s using small samples. Subsequently, these efforts were expanded to larger arrays in experiments such as MiDBD, Cuoricino, and CUORE-0. Currently, this research has reached its pinnacle with the CUORE ton-scale experiment \cite{Adams:2022a}, which is operating at the Gran Sasso underground laboratory (LNGS) in Italy. 
(The evolution of TeO$_2$-based bolometric $2\beta$ search experiments is overviewed in \cite{Brofferio:2018}.)

All past and present TeO$_2$-based $2\beta$ experiments exploited and currently used crystals with natural Te content, with the exception of two samples enriched in $^{130}$Te used in addition to natural ones in the MiDBD \cite{Pirro:2000} and Cuoricino \cite{Bryant:2010} experiments. Due to poor performance and radiopurity of enriched samples compared to natural crystals, as well as taking into account the still high natural isotopic abundance of $^{130}$Te, $^{130}$Te-enriched crystals were abandoned from the CUORE program. However, the interest in $^{130}$TeO$_2$ has raised again in view of the CUORE upgrade \cite{Artusa:2014wnl}, particularly in the CUPID program \cite{Wang:2015raa,Wang:2015taa}. Two new enriched samples have been produced and tested at LNGS showing high performance and low radioactivity \cite{Artusa:2017}. Due to challenging particle identification with TeO$_2$ bolometers \cite{Poda:2017}, mainly exploiting the detection of $\gamma(\beta)$-induced Cherenkov radiation proposed in \cite{Tabarelli:2010}, and difficulties in further reduction of $\gamma$ background in the CUORE cryostat, CUPID has selected lithium molybdate (Li$_2$MoO$_4$) scintillating crystals \cite{CUPIDInterestGroup:2019inu}. This compound allows for a more efficient scintillation-assisted particle identification and a lower $\gamma$ background, by moving the region of interest ($Q_{2\beta}$ of $^{100}$Mo is 3034 keV \cite{Wang:2021a}) above the end-point of the most intense natural $\gamma$ radioactivity (2615~keV $\gamma$ of  $^{208}$Tl from the $^{232}$Th family).

Nevertheless, $^{130}$Te remains one of a few $2\beta$-active isotopes highly suitable for multi-isotope bolometric study \cite{Giuliani:2018}. Furthermore, TeO$_2$ material has a few advantages compared with Li$_2$MoO$_4$: 
i) higher detection efficiency for the same volume (higher density and effective atomic number); 
ii) no hygroscopic properties; 
iii) two orders of magnitude longer half-life of the ordinary $2\nu2\beta$ decay, which makes the background contribution from random coincidences of the associated events negligible for $^{130}$Te-enriched thermal detectors, whereas this is the major component of background expected in CUPID with $^{100}$Mo-enriched scintillating bolometers \cite{Chernyak:2012}. 
Taking into account the possibility to improve the present, already extremely low, $\gamma$ background in CUORE by even more careful selection of external screens with lower contamination and/or by applying innovative techniques for active suppression of backgrounds induced by $\gamma$ (e.g. proposed by BINGO \cite{Armatol:2023}) and $\alpha$ (e.g. by detection of $\gamma$($\beta$)-induced Cherenkov radiation from TeO$_2$ \cite{Tabarelli:2010}), the choice of $^{130}$TeO$_2$ for future high-sensitivity $2\beta$ searches is still attractive and viable.

With this in mind, TeO$_2$ together with Li$_2$MoO$_4$ detector materials have been chosen for the implementation of the CROSS program on the development of metal-coated bolometers with the capability of pulse-shape discrimination between bulk and near surface interactions \cite{Bandac:2020,Bandac:2021}. The validation of the CROSS technology for active rejection of surface-originated $\alpha$ and $\beta$ backgrounds is planned to be realized in a small-scale demonstrator with tens of Li$_2$$^{100}$MoO$_4$ and a few $^{130}$TeO$_2$ bolometers. 
For the sake of completeness, we refer to early attempts in the development of active background rejection for bolometric $2\beta$ detectors, including TeO$_2$, which are summarized in \cite{Poda:2017} and we underline a couple of currently ongoing projects on this topic. 
As in CROSS, TeO$_2$ and Li$_2$MoO$_4$ are both included in the  BINGO project, which is developing an innovative detector assembly system with high-performance bolometric photodetectors and a cryogenic active veto based on bismuth germanate crystals with bolometric photodetection \cite{Armatol:2023}. Recently, the BINGO detector system with the polyamide wire serving to hold a similar size Li$_2$MoO$_4$ crystal (twice lighter than TeO$_2$) have been tested underground in the CROSS cryostat and good detector performance has been reported \cite{Armatol:2024}. 
Also, the idea of scintillating polymer-based holder allowing active background rejection of radioactivity induced by close materials of the detector structure has been proposed \cite{Biassoni:2023}. First prototypes with small and large TeO$_2$ bolometers were tested in such a 3D-printed detector holder \cite{Biassoni:2021,Biassoni:2023}, as a part of R\&D for next-generation $2\beta$ experiments with this detector material.

Therefore, interest in TeO$_2$ low-temperature detectors is still high for $2\beta$ search applications. With this aim, we report on the development of large-volume $^{130}$TeO$_2$ crystals for the CROSS experiment. We present then the first results of bolometric measurements both with newly developed small-size natural crystals and large-size $^{130}$Te-enriched crystals, operated in dilution refrigerators of above-ground and underground laboratories respectively.

\section{Development of TeO$_2$ crystals from $^{130}$Te-enriched tellurium}

\subsection{Initial $^{130}$Te-enriched material}
\label{sec:Initial_Te}

The initial tellurium powder enriched in $^{130}$Te isotope to around 93\% (provided by the University of South Carolina) was assayed as received from the customer for impurities and oxygen content. Using the inductively coupled plasma mass-spectrometry (ICP-MS), the total concentration of impurities excluding oxygen was measured at 25~ppm; with the duplicate sample, a consistent result of 27~ppm was obtained. The detected impurities are listed in table \ref{tab:Te_contaminants}. The oxygen content is found on the level of 2\%.

\begin{table}
\centering
\caption{ICP-MS-detected impurities in samples which were produced during an R\&D on purification of the $^{130}$Te-enriched material. We quote results for initial (as-purchased) raw material, after purification (average for pure section of the ingot produced by solidification, see text), and after the conversion to a $^{130}$TeO$_2$ powder. Uncertainties are at 20\% level. 
No results are given for cases when the signal is below the detection limit (including the following elements: Ag, As, B, Bi, Cd, Co, Mo, P, Sb, Se, and V).}
\smallskip
\begin{tabular}{cccc}
\hline
Element  & \multicolumn{3}{c}{Concentration (ppm) of element in $^{130}$Te-enriched samples}   \\
~ & Raw material & After solidification & $^{130}$TeO$_2$ powder \\
\hline
\hline
Al  & 0.6  &  & 0.4       \\
Ca  & 1.0 & 0.3 & 0.3       \\
Cr  & 0.6  & 0.2 & 0.5       \\
Cu  & 2.8  &  &  0.3      \\
Fe  & 8.6  & 1.8 & 1.7       \\
K   & 0.4  &  &        \\
Li  & 0.1  &  &        \\
Mg  & 2.5  & 0.2 &        \\
Mn  & 0.1  &  &        \\
Na  & 0.4  &  & 0.2       \\
Ni  & 1.4  & 0.1 & 0.3       \\
Si  & 2.7  &  & 1.6       \\
Sn  &   &  &  0.2      \\
Ti  & 0.1  &  &        \\
Zn  & 3.1  & 0.5 & 0.2       \\
\hline
Total   & 24.6  & 3.1 & 5.7       \\
\hline
\end{tabular}
\label{tab:Te_contaminants}
\end{table}

\subsection{Purification by the solidification method}

\begin{figure}
\centering
\includegraphics[width=0.7\textwidth]{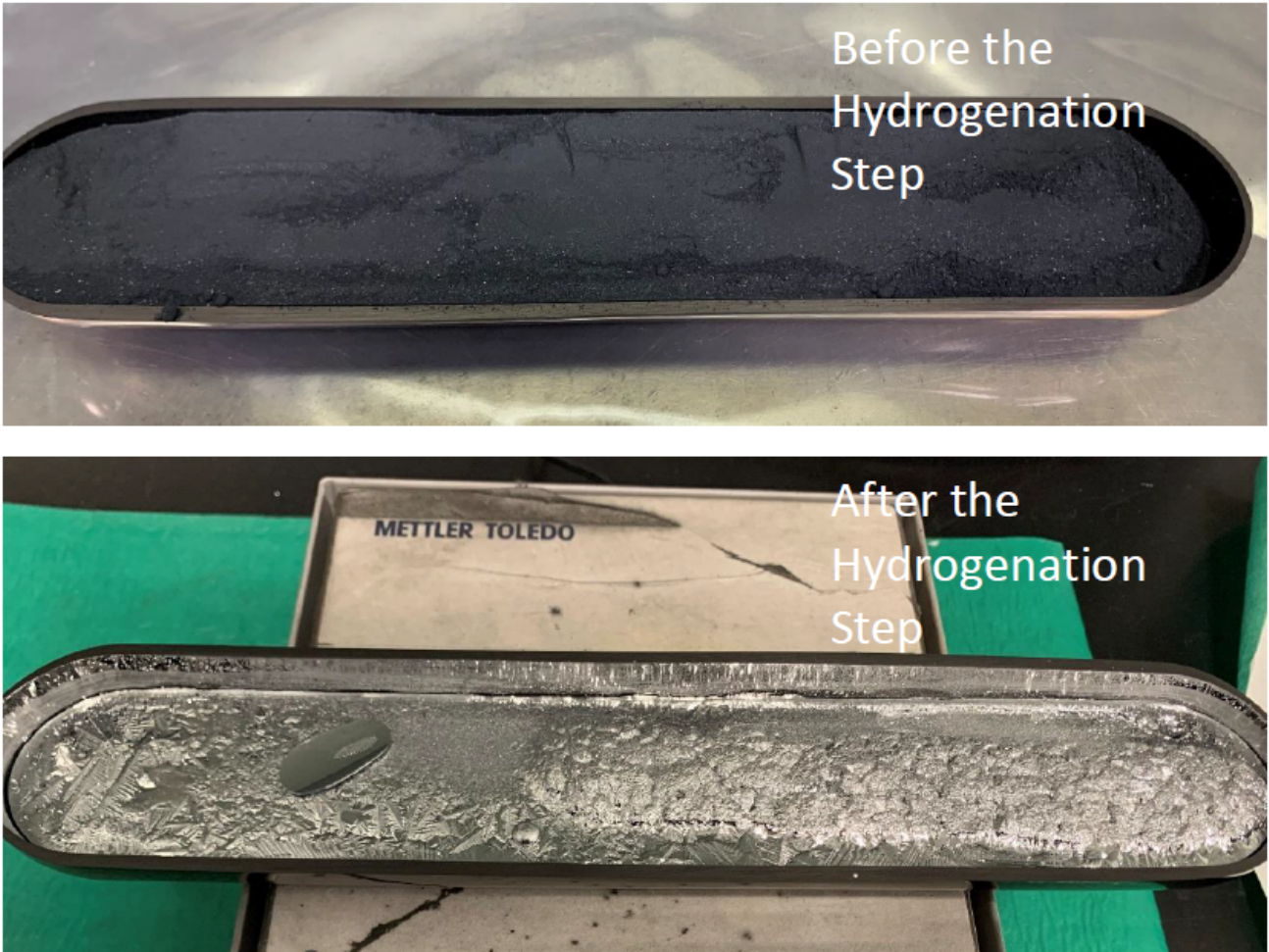}
\caption{View of a $^{130}$Te-enriched sample before (top) and after (bottom) the hydrogenation step. A 1 kg load of the material is used for this demonstration test.}
\label{fig:Te_hydrogenation}
\end{figure}

The purification of the enriched material has been performed in Canada (5N company, Garand, Montreal, Quebec). At the initial step --- hydrogenation process \cite{mimura2002hydrogen} --- a deoxidized ingot was produced from the tellurium oxide powder. The material was remelted under a stream of hydrogen to obtain a solid ingot, as illustrated in figure \ref{fig:Te_hydrogenation}.

\begin{figure}
\centering
\includegraphics[width=0.45\textwidth]{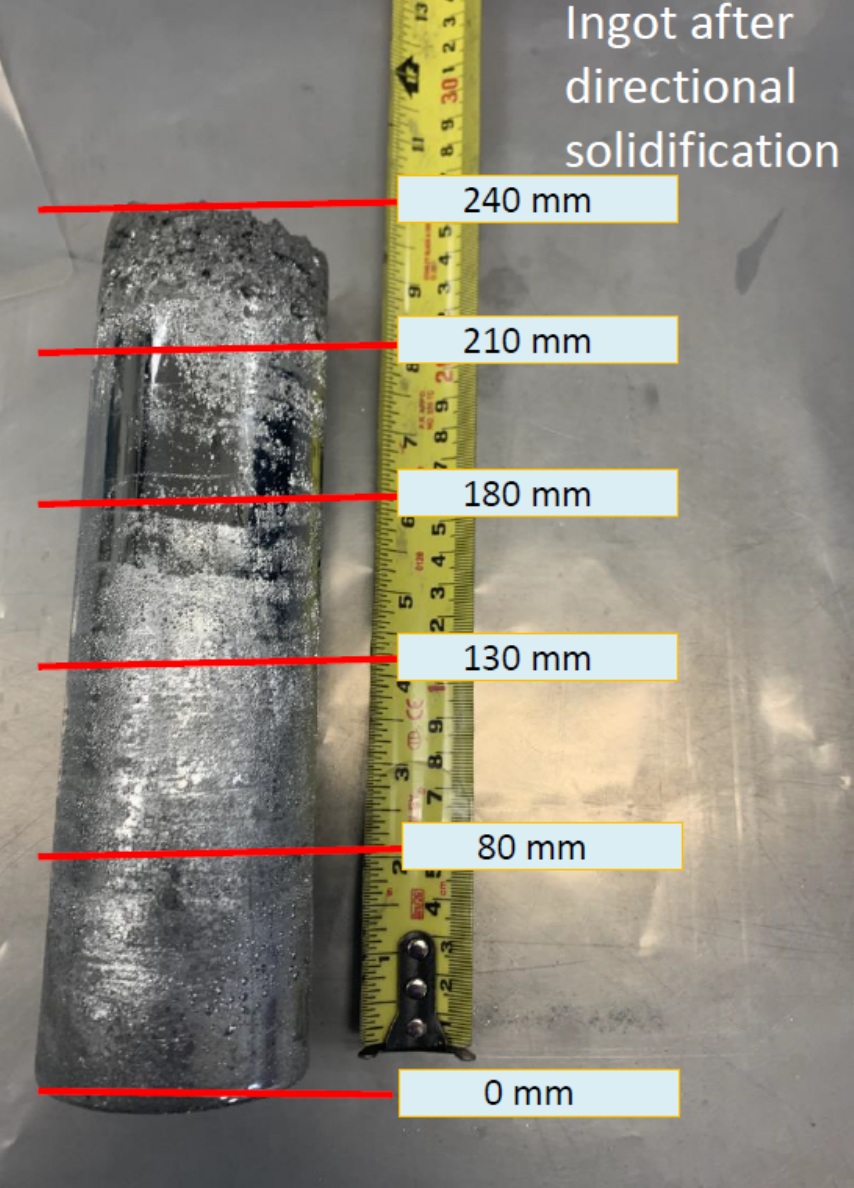}
\caption{View of the $^{130}$Te-enriched tellurium ingot produced by the directional solidification method. Red lines show the positions of the probes (in the center and on the surface) used in the study of segregation profiles of the detected impurities, exploited to define the poorest part of the present ingot (below 180 mm, see details in text).}
\label{fig:Te_solidification}
\end{figure}

\begin{figure}
\centering
\includegraphics[width=0.7\textwidth]{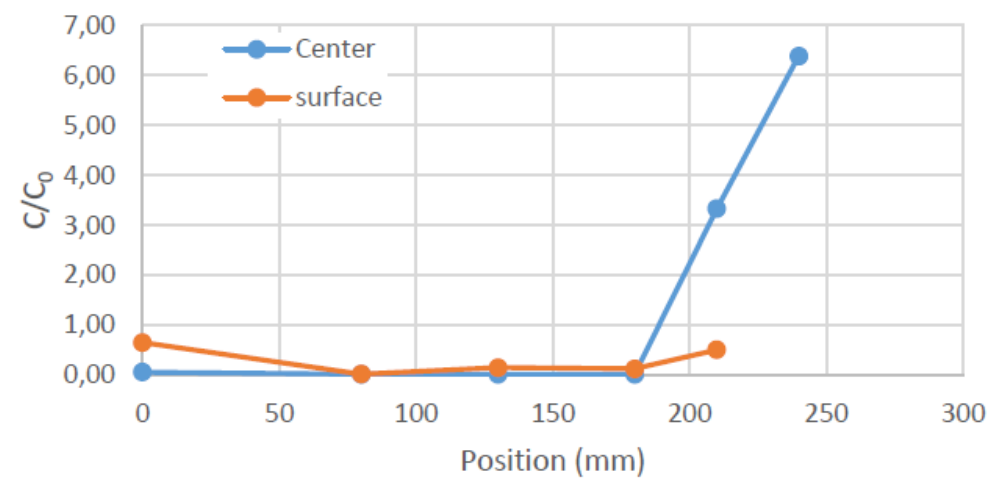}
\caption{Segregation profile of iron (the main impurity) in samples taken from the center (blue) and on the surface (orange) of the $^{130}$Te-enriched tellurium ingot produced by the directional solidification.}
\label{fig:Fe_segregation_profile}
\end{figure}

Then the directional solidification method \cite{kekesi2002principles} was applied. The tellurium ingot from the hydrogenation step was remelted under a partial atmosphere of hydrogen and solidified directionally to get impurity segregation and to obtain a material with a 5N purity. After the process, the ingot was cut at 6 positions over the length, as shown in figure \ref{fig:Te_solidification}, and each position was sampled in the center and on the surface of the ingot. Each sample from the different positions in the ingot was assayed to get segregation profiles, showing signals of impurities above the detection limit at positions of 180--210 mm. This is illustrated in figure \ref{fig:Fe_segregation_profile} for the main impurity (Fe) of the $^{130}$Te-enriched material; the segregation profiles of other impurities show the same trend. Therefore, the impure section above 180 mm was removed, while the rest of the ingot was ground and the powder was homogenized for a final sampling. 
The final assay of the purified tellurium, done using ICP-MS measurements, is presented in table \ref{tab:Te_contaminants}. The total impurity in the $^{130}$Te-enriched powder is detected at 3 ppm, showing no hint of a re-contamination of the material to be used for the crystal growth.

For the first run of purification a yield of 68\% was achieved. This figure has been improved to $\sim$80\% in the following batch (of 7 kg) with further improvement of the deoxidation step and a better definition of the impure section after the directional solidification.  
A further purification of the rejected material is considered later in the program.

\subsection{Conversion to $^{130}$TeO$_2$ powder}
 
\begin{figure}
\centering
\includegraphics[width=0.5\textwidth]{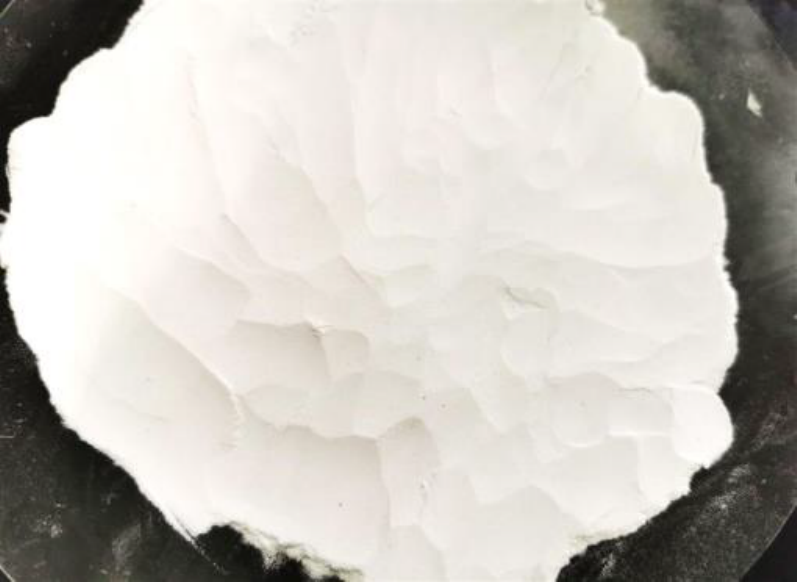}
\caption{TeO$_2$ powder made of highly purified tellurium enriched in $^{130}$Te to 93\%.}
\label{fig:TeO_powder}
\end{figure}

Tellurium oxidation assay was also performed by the 5N company using the material purified via the directional solidification and following a standard process. A 200 g load of tellurium was ground into a powder to carry out the oxidation test. The obtained TeO$_2$ powder was found visually white (see in figure \ref{fig:TeO_powder}) and differential scanning calorimetry assay revealed no unoxidized tellurium (detection limit was 100 ppm). A particles size distribution analysis was performed on the final powder using laser scattering particle size  analyzer (Horiba, LA-950); the mean particle size is measured as 43 $\mu$m.  Impurity levels investigated with ICP-MS are presented in table \ref{tab:Te_contaminants}. The total level of impurities is low and meets the objective of 10 ppm limit. The main impurities are iron at 1.7 ppm and silicon at 1.6 ppm.

Thus, all the purified material has been oxidized using the standard process. The yield is close to 80\% and the total purification factor is about 10, with a residual average impurity concentration of 1.4 ppm in the final purified $^{130}$TeO$_2$ powder.

\subsection{Crystal growth}

We identified a company in the US, Gooch \& Housego (Clevelend, Ohio), for the Czochralski growth of large-volume (about 90 cm$^3$) $^{130}$Te-enriched TeO$_2$ crystals to be included in the CROSS demonstrator. In order to validate the crystal growth process, two natural TeO$_2$ samples (4~cm$^3$ each) were produced by this company from non-purified materials and tested as Cherenkov-radiation-emitting bolometers in Orsay to demonstrate acceptable bolometric performance and the level of internal $^{210}$Pb contamination (less than 10 mBq/kg), as detailed in the next section. After the validation of the TeO$_2$ crystal producer, we proceeded with the growth of $^{130}$Te-enriched crystals and six $^{130}$TeO$_2$ samples 45 $\times$ 45 $\times$ 45~mm each were fabricated using purified starting materials. 

In order to investigate the isotopic composition of Te in the grown $^{130}$TeO$_2$ crystals, we took fragments (about 80 mg in total) of crystals coming from different locations in the ingot, dissolved them in a mixture of HCl and HNO$_3$ acids, and carried out the ICP-MS measurement in isotopic mode. The results on the Te isotopic concentration in the $^{130}$Te-enriched crystals are given in table \ref{tab:Te_isotopes}. The enrichment in the isotope tellurium-130 is found at $\sim$91\%, which is still very high compared to the enrichment of the initial powder ($\sim$93\%). Thus, the adoption of the Czochralski technique (i.e. a small seed) avoids an important dilution of the enriched material in the growth seed. This result would be impossible using the Bridgman method, e.g. employed for the production of the natural TeO$_2$ crystals for the CUORE experiment, as in this case the seed is very large and dominate the final crystal composition. This result is innovative and very important for future bolometric $2\beta$ decay search experiments based on enriched tellurium.

\begin{table}[h]
\centering
\caption{ICP-MS-measured isotopic composition of tellurium in $^{130}$Te-enriched crystals developed in the present work. Isotopic abundance of natural tellurium is given for comparison.}
\smallskip
\begin{tabular}{ccc}
\hline
Te isotopes  & \multicolumn{2}{c}{Isotopic abundance (\%) in tellurium samples} \\
~ & Enriched in $^{130}$Te  & Natural Te \cite{Meija:2016}  \\
\hline
\hline
$^{120}$Te         &  0.00020(4) & 0.0009(1)      \\
$^{122}$Te         &  0.0038(8) & 0.0255(12)       \\
$^{123}$Te         &  0.013(3) & 0.0089(3)       \\
$^{124}$Te         &  0.072(15) & 0.0474(14)       \\
$^{125}$Te         &  0.11(3) & 0.0707(15)       \\
$^{126}$Te         &  0.30(5) & 0.1884(25)       \\
$^{128}$Te         &  8.1(2) & 0.3174(8)       \\
$^{130}$Te         & 91.4(5) & 0.3408(62)       \\
\hline
\end{tabular}
\label{tab:Te_isotopes}
\end{table}

\section{Operation of TeO$_2$-based low-temperature detectors}

\subsection{Above-ground test of small-size natural precursors}
\label{sec:test_ijclab}

\begin{figure}
\centering
\includegraphics[width=0.8\textwidth]{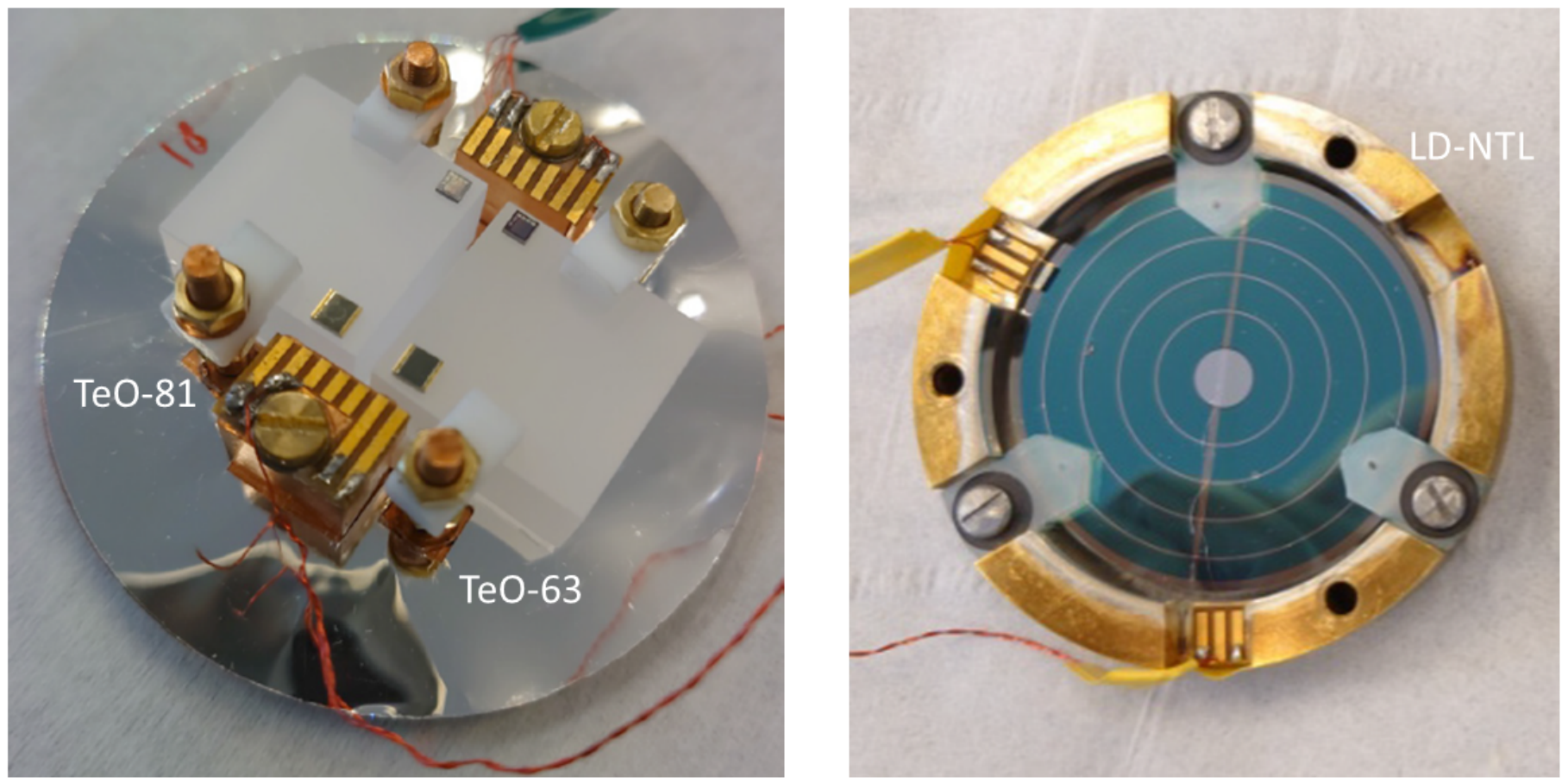}
\caption{Bolometers based on two 20 $\times$ 20 $\times$ 10-mm TeO$_2$ crystals (Left) and a $\oslash$44-mm Ge slab with a 0.2~mm thickness (Right, from \cite{ZnO_bolo:2023}), used in the present work for the detection of Cherenkov radiation induced by particle interactions in the TeO$_2$ thermal detectors.}
\label{fig:TeO_natural_detector}
\end{figure}

In order to validate the selected producer of $^{130}$TeO$_2$ crystals for CROSS, we carried out a low-temperature characterization of two TeO$_2$ small-size samples (20 $\times$ 20 $\times$ 10 mm, 24 g each), grown from high-purity powder with natural Te isotopic composition (without the additional purification foreseen in CROSS). The detector construction was carried out in the same way as CROSS prototypes used in studies of metal-coated bolometers \cite{Bandac:2020}. Specifically, each crystal was equipped with a temperature sensor (NTD-Ge, neutron-transmutation-doped Ge \cite{Haller:1994}) and a heater (heavily doped Si chip \cite{Andreotti:2012}); both elements were glued with a bi-component epoxy (Araldite\textregistered) on a 20 $\times$ 20-mm side of a TeO$_2$ crystal. The crystals were mounted on a Cu plate with the help of two Cu rods and PTFE (polytetrafluoroethylene) spacers. The assembly was secured by nuts. To provide electrical connections, we wire-bonded sensors and heaters with Au wires to the gold-plated-on-Kapton contacts. The assembled TeO$_2$ bolometric module is shown in figure \ref{fig:TeO_natural_detector} (Left). The lateral part of the assembly was covered by a Cu cylinder; the internal side of this screen and the Cu plate facing TeO$_2$ crystals were covered by a reflective film (Vikuiti\texttrademark).

On the top of the TeO$_2$ assembly we placed a low-threshold optical bolometer \cite{Novati:2019} to attempt to detect the tiny Cherenkov radiation induced by $\beta$ / $\gamma$ / $\mu$ interactions in the TeO$_2$ bolometers \cite{Berge:2018}, thus allowing for selection and study of $\alpha$ events. This photodetector (NTLLD2 in \cite{Novati:2019}) was made of an electronic grade pure Ge slab ($\oslash44\times0.18$~mm) and instrumented with a smaller size NTD-Ge sensor epoxy-glued on the wafer surface. The Ge wafer was coated by a 70-nm SiO layer to reduce the reflectivity of the surface and thus to improve photon collection. A set of aluminum concentric electrodes were deposited on the Ge surface and interconnected with Al wires, allowing to apply a voltage bias and to amplify thermal signals exploiting the Neganov-Trofimov-Luke effect in semiconductors \cite{Neganov:1985, Luke:1988}. The Ge slab was kept in a Cu housing with the help of sapphire balls ($\varnothing 1.5$~mm) and polypropylene supporting elements; the NTD-Ge sensor was polarized through Au bonding wires. The mounted optical bolometer is depicted in figure \ref{fig:TeO_natural_detector} (Right). The photodetector was facing the TeO$_2$ crystals by the side carrying the Al electrodes (to get more efficient amplification of photon-induced thermal signals), while the opposite side was facing another bolometer (ZnO). Its recently published characterization \cite{ZnO_bolo:2023} contains more details on the experiment and results of the light detector performance, which are only briefly described below.

The fully assembled device was tested with the help of a pulse-tube-based dilution refrigerator \cite{Mancuso:2014} placed at IJCLab (Orsay, France) inside a 10-cm-thick lead shielding. The detector module was spring-suspended inside the cryostat, having a flexible thermal link to the coldest part of the unit to mitigate vibrational noise induced by the pulse tube and pumps \cite{Olivieri:2017}. The detectors were cooled down to 15 mK. The NTD-Ge sensors were polarized with $O$(1 nA) currents to get  working resistances below 10 M$\Omega$, in compliance with the electronics used (e.g. the set of the load resistances) and to avoid a large contribution of the Johnson noise. The continuous streams of data were acquired by 16-bit National Instruments cards (NI USB-6218 BNC) with a sampling rate of 5~kS/s; an anti-aliasing Bessel filter with a  cut-off frequency of 675~Hz was used in the readout. The data processing was realized using an optimum filter technique \cite{Gatti:1986} implemented in a MATLAB-based application \cite{Mancuso:2016} with data-driven templates of signal and noise of the detectors used to construct the transfer function of the filter. 

\begin{figure}
\centering
\includegraphics[width=0.7\textwidth]{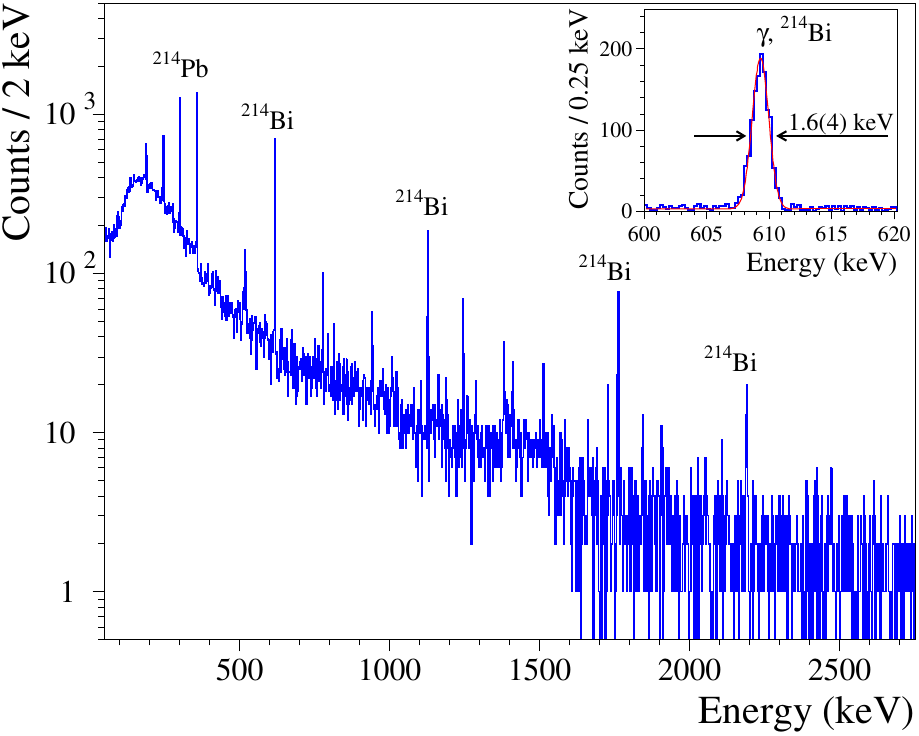}
\caption{A partial energy spectrum measured by a 24-g TeO$_2$ bolometer in a low-temperature test (15~mK; 96 h of data taking) in an above-ground laboratory. The most intense $\gamma$ peaks, corresponding to $^{214}$Pb and $^{214}$Bi, are labeled. A 609-keV peak of $^{214}$Bi together with a fit to a Gaussian plus a flat background component are shown in the Inset. The energy resolution of the peak is measured as 1.6(4) keV FWHM.}
\label{fig:TeO_natural_Bkg_spectrum}
\end{figure}

Measurements were carried out with a $^{232}$Th calibration source for 24 h and without the use of additional radioactive sources; the nominally background data were taken for a total of 166 hours. Due to a comparatively high environmental $\gamma$-radioactivity (mainly from decays of the $^{226}$Ra sub-chain \cite{Bekker:2016}), and thanks to high spectroscopic properties of the TeO$_2$ bolometers, background data can be easily calibrated because of the presence of many sharp peaks in a wide energy range as seen in figure \ref{fig:TeO_natural_Bkg_spectrum}. We observed that both TeO$_2$ thermal detectors show excellent energy resolution, comparable to results achieved with bare (i.e. non-coated) samples of the same-size used in the CROSS program \cite{Bandac:2020,Khalife:2021}. 
The detectors are characterized by high sensitivity defined as the voltage signal per unit of deposited energy  (0.2--0.5 $\mu$V/keV) and a low noise represented by the baseline width at the optimum filter output (0.25--0.60 eV RMS). As for the time properties of the thermal response, both TeO$_2$ bolometers show typically slow signals with the ascending part of about 12 ms and with the descending part about 60 ms. The performances of the TeO$_2$ thermal detectors are summarized in table \ref{tab:TeO_natural_performance}.

\begin{table}
\centering
\caption{Performance of two 24-g TeO$_2$ bolometers and a Ge light detector exploiting the signal amplification based on the Neganov-Trofimov-Luke effect (60 V bias on Al electrode). We report the working point of the NTD thermistor (resistance $R_{NTD}$ at a given current $I_{NTD}$), pulse-shape time constants (rise- and decay-time parameters, $\tau_{r}$ and $\tau_{d}$), detector sensitivity represented by the signal voltage amplitude per unit of the deposited energy ($A_s$; estimated with 609-keV $\gamma$ quanta), and RMS baseline width (RMS$_{noise}$).}
\smallskip
\begin{tabular}{ccccccc}
\hline
Detector  & $R_{NTD}$ & $I_{NTD}$ & $\tau_{r}$ & $\tau_{d}$ & $A_s$ & RMS$_{noise}$ \\
ID & (M$\Omega$)  & (nA) & (ms) &  (ms) & ($\mu$V/keV) & (keV)  \\
\hline
\hline
TeO-63         & 8.8 & 1.3 & 12.7  & 67  & 0.47    & 0.59       \\
TeO-81         & 5.0 & 1.5 & 12.4  & 61  & 0.22    & 0.25       \\
\hline
LD-NTL \cite{ZnO_bolo:2023}         & 4.2 & 1.3 & 1.3   & 11  & 38     & 0.017       \\
\hline
\end{tabular}
\label{tab:TeO_natural_performance}
\end{table}

\begin{figure}
\centering
\includegraphics[width=0.7\textwidth]{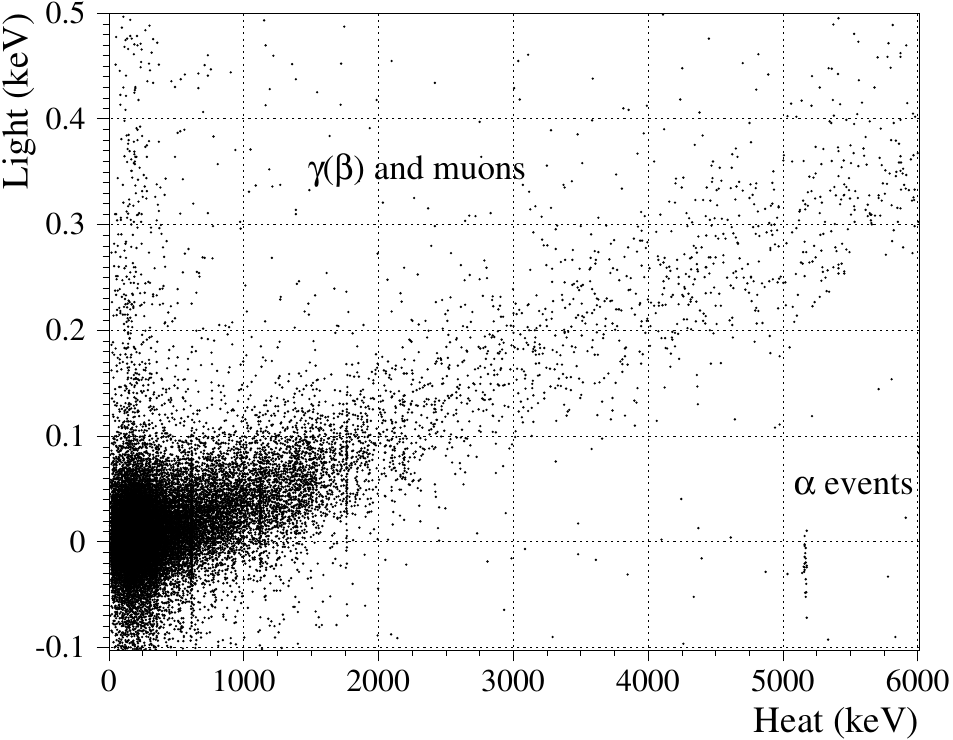}
\caption{Light signals detected by the Ge optical bolometer versus heat signals measured by the 24-g TeO$_2$ bolometer (TeO-81) over 96 h of exposure to natural radioactivity at a sea-level laboratory. The major part of events below $\sim$2.6 MeV corresponds to $\gamma$($\beta$) interactions in the crystal medium, while muon-induced events represent dominant contribution above this energy. A structure of high-energy TeO$_2$-detected events which exhibit no light signal correspond to $\alpha$ decays of $^{210}$Po inside the TeO$_2$ detector medium, indicating an internal contamination by $^{210}$Po($^{210}$Pb). The energy scale of the detector, calibrated using 609-keV $\gamma$ quanta of $^{214}$Bi, exhibits an evident non-linearity, which results to a few \% miscalibration of the 5407-keV $\alpha$ peak of $^{210}$Po (see discussion in Sec. \ref{sec:test_lsc}).}
\label{fig:TeO_natural_Light_Heat}
\end{figure}

During the TeO$_2$ detector calibrations we kept the Al electrodes of the Ge bolometer unpolarized, thus allowing the detection of Te X-rays induced by environmental radioactivity in the crystals, which was used to calibrate the maximum distribution of the muon-induced events \cite{ZnO_bolo:2023}. Knowing the energy scale of the Ge bolometer, we found a factor $\sim$10 improvement of the signal-to-noise ratio by applying a 60 V bias on the Al electrode, resulting to a high performance reported in table \ref{tab:TeO_natural_performance} and in \cite{ZnO_bolo:2023}. 

Thanks to the low noise achieved with the Ge bolometer, we were able to detect the TeO$_2$-emitted Cherenkov light, providing highly efficient particle identification as illustrated in figure \ref{fig:TeO_natural_Light_Heat}. As one can see in figure \ref{fig:TeO_natural_Light_Heat}, most of the events populate a band including $\beta$-, $\gamma$- and $\mu$-induced interactions, which are associated with a registered Cherenkov light of the order of 40 eV per MeV energy deposited in the TeO$_2$ bolometer. 
The population of $\alpha$ events, which do not produce the Cherenkov radiation (due to kinematic energy threshold of 400 MeV), is also observed in the data (see figure \ref{fig:TeO_natural_Light_Heat}). Moreover, these $\alpha$-events have even negative values of light signals due to a small cross-talk, also seen in the data for heater-induced events (not shown in figure \ref{fig:TeO_natural_Light_Heat}). 

The observed $\alpha$ events are mainly ascribed to decays of $^{210}$Po ($Q_{\alpha}$ = 5407.5~keV, $T_{1/2}$ = 138.376~d) from $^{210}$Po/$^{210}$Pb contaminations in the crystal bulk corresponding to an activity of 2.8(4) mBq/kg and 9.1(8) mBq/kg in the samples TeO-63 and TeO-81 respectively. The $^{210}$Po $\alpha$ activity is typical for different scintillators \cite{Danevich:2018} and mostly related to the $^{210}$Pb content in materials. In case of TeO$_2$ crystals, the chemical affinity between Te and Po could result to a less efficient segregation of Po isotopes, including $^{210}$Po, during the TeO$_2$ powder purification and crystal growth.  
It is worth noting that we detected only tens of $^{210}$Po events, while the presence of other $\alpha$-active radionuclides from U/Th chains is not evident, and thus their activities are expected to be at least an order of magnitude lower. Therefore, one can conclude that the radiopurity of both samples is very high, especially keeping in mind that the starting material did not pass an extra purification.

Accordingly, the results of the low-temperature test presented above prove high bolometric and spectrometric properties together with a low level of radioactive contamination of the TeO$_2$ crystals, and validate the crystal producer for the development of the $^{130}$Te-enriched crystals for the CROSS experiment.

\subsection{Underground test of large-size $^{130}$Te-enriched crystals}
\label{sec:test_lsc}

\begin{figure}
\centering
\includegraphics[width=0.9\textwidth]{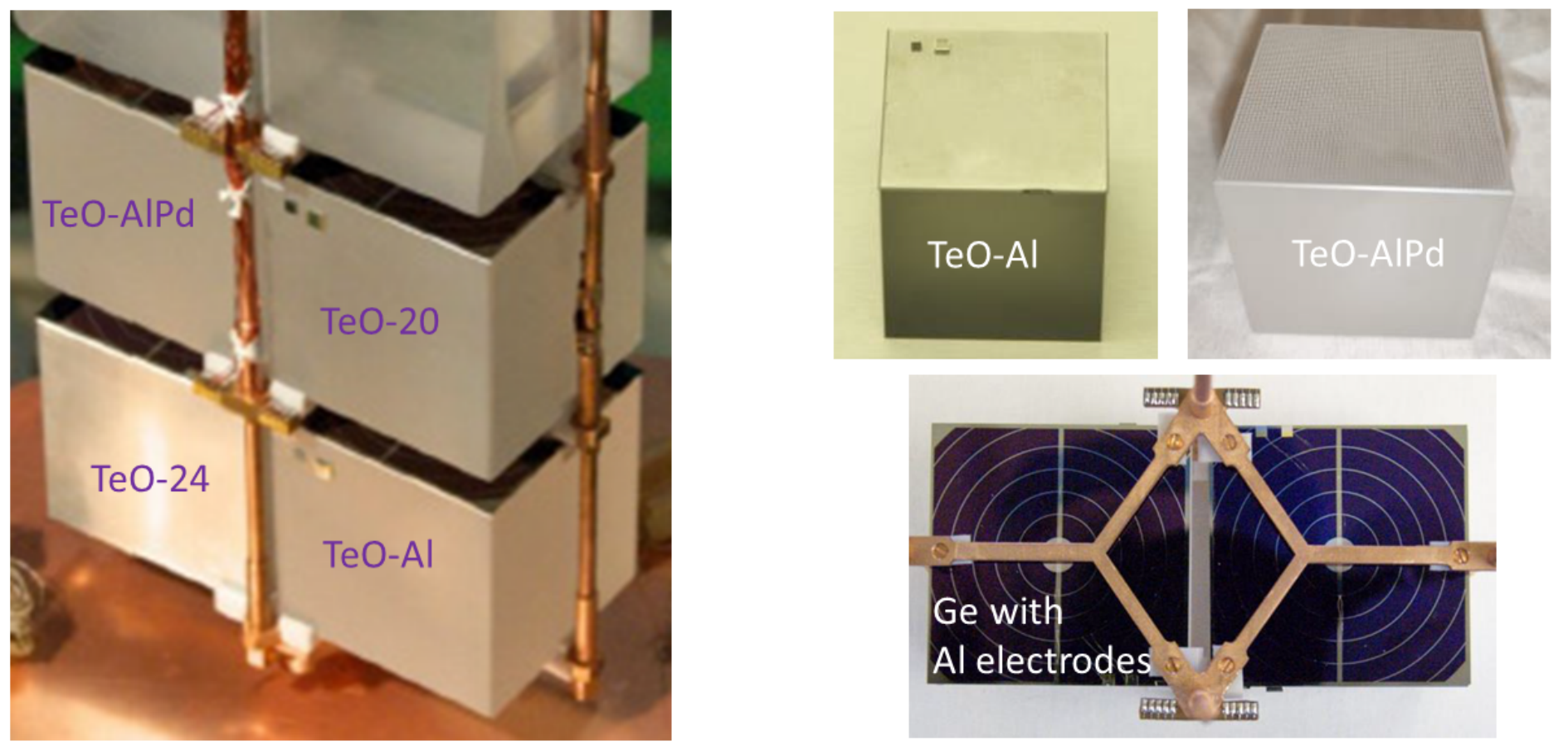}
\caption{Fragment of a 5-floor array of dual-readout bolometers showing the last two floors with large-volume $^{130}$TeO$_2$ bolometers (Left). Two $^{130}$TeO$_2$ crystals are coated with Al film on 4 sides and AlPd grid on a single side respectively (Right top). Thin square-shaped Ge bolometers with concentric Al electrodes (light detectors with Neganov-Trofimov-Luke effect signal amplification) are placed above the $^{130}$TeO$_2$ thermal detectors (Right bottom).}
\label{fig:TeO_enriched_detectors}
\end{figure}

To test the newly developed $^{130}$Te-enriched crystals as bolometers, we took randomly four of the six cubic samples (45~mm side each) produced for the CROSS project. Two chosen samples (TeO-20 and TeO-24) were used uncoated, while the other two crystals were metal-coated as follows: i) four-side coated with a 200~nm Al layer (TeO-Al); ii) one-side coated with an AlPd bi-layer, 100~nm / 10~nm, in the form of a grid (TeO-AlPd). Each crystal was equipped with an NTD-Ge thermistor and a Si:P heater, coupled with UV-curable (PERMABOND 620) and epoxy (Araldite\textregistered ~Rapid) glue respectively. The samples were assembled in the Cu structure with 3D-printed PLA (polylactic acid) spacers --- the holder \emph{Slim} version designed for the CROSS experiment \cite{CROSSdetectorStructure:2024}--- fasting two crystals per floor. In addition, a thin Ge bolometer (45 $\times$ 45 $\times$ 0.3~mm) has been mounted on each TeO$_2$ crystal to detect Cherenkov radiation. All Ge slabs were SiO-coated (70 nm) to reduce the reflectivity of the surface. A set of Al concentric electrodes, not optimized for the adopted squared geometry of the light detectors, was deposited following the procedure described in \cite{Novati:2019}. The purpose of the electrodes is to amplify the thermal signal via the Neganov-Trofimov-Luke effect, as used in the above presented measurements (see section \ref{sec:test_ijclab}). In total, the tower was composed of 5 floors; the rest of the array was made of six Li$_2$MoO$_4$-based bolometers and the same type of Ge light detectors attached to them \cite{CUPIDalternativeStructure:2024}. The part of the tower containing the $^{130}$TeO$_2$ bolometers and construction elements of the detector structure are depicted in figure \ref{fig:TeO_enriched_detectors}.

The detector array was installed inside a low-background facility with a pulse-tube-based dilution refrigerator, built in the Canfranc underground laboratory (Spain) for the CROSS experiment \cite{Olivieri:2020,Armatol:2021b}. In order to mitigate the vibrational noise induced by the pulse tube cryocooler, and to gain space of the available experimental volume, the set-up was recently upgraded with a pendulum-like detector suspension with three magnetic dampers installed at the 1 K stage \cite{CROSS_Magnetic_dampers:2023}. Thanks to the underground location of the facility and to massive external and internal shields, the background level from natural radioactivity is well reduced providing suitable conditions for characterisation of large-volume low-temperature detectors \cite{CROSS_Magnetic_dampers:2023,CROSSdeplLMO:2023}. After reaching a base temperature of around 12 mK, we carried out the characterization of the bolometers at warmer temperatures (17, 22, and 27 mK) regulated on the detector plate. The detector array was controlled and read-out using room-temperature electronics \cite{Carniti:2020,Carniti:2023}. The amplified voltage output of the NTD-Ge thermistors was digitized in continuous streams by a 24-bit ADC with a sampling rate of 2 or 10 kHz; the cut-off frequency of the active Bessel filter was set as 0.3 or 2.5 kHz respectively. The optimum-filter-based data processing was done with the same tool, and in a similar way, as described above (section \ref{sec:test_ijclab}).

At low temperature, we lost no contacts but we faced an issue with the readout of a single $^{130}$TeO$_2$ channel (TeO-24), therefore we grounded it during the experiment. A similar issue happened for another channel (TeO-Al) after the electronics upgrade \footnote{Currently, 11 over 14 front-end boxes and twelve 12-channel DAQ boards have been installed.}, thus we were not able to study this detector at the warmest temperature (27 mK). 

\begin{figure}
\centering
\includegraphics[width=0.49\textwidth]{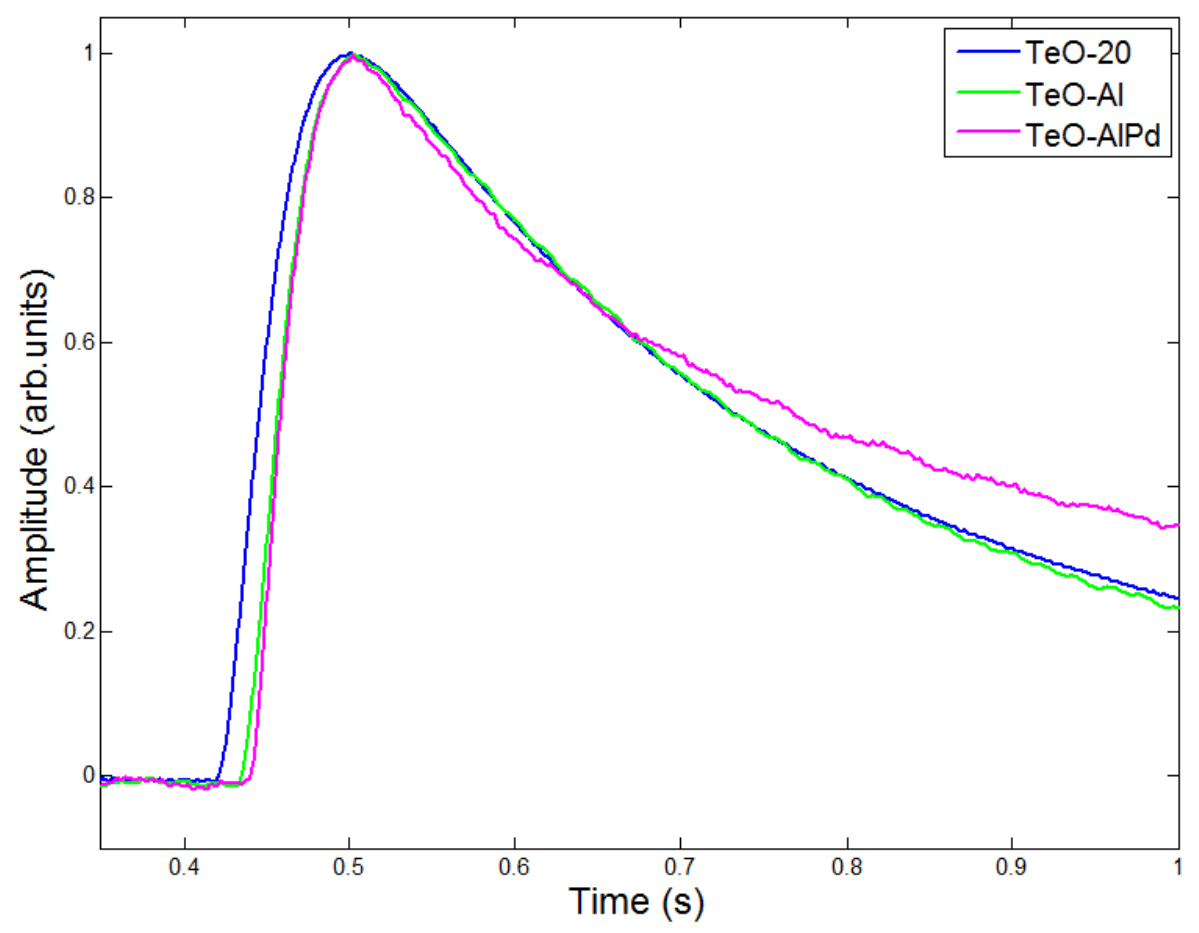}
\includegraphics[width=0.49\textwidth]{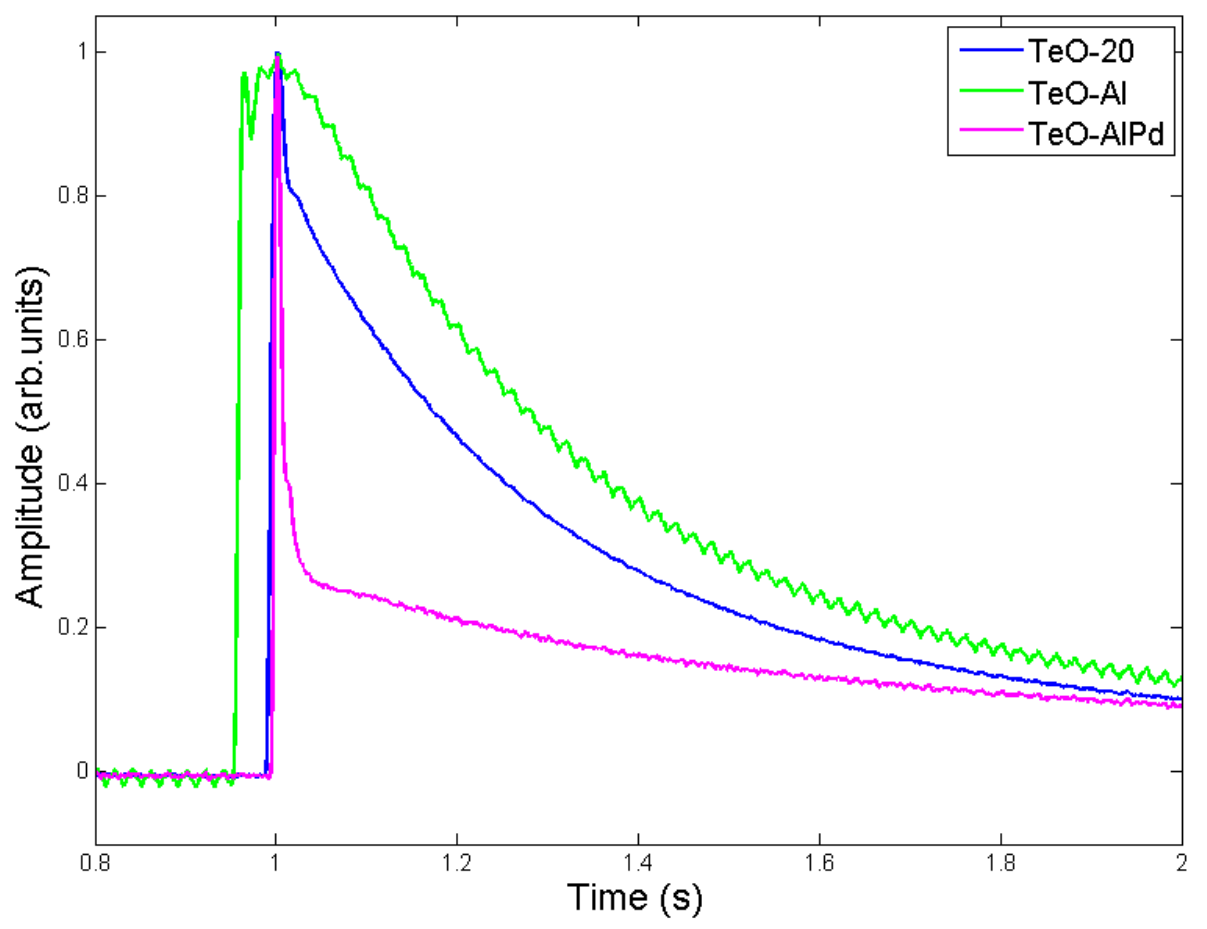}
\includegraphics[width=0.49\textwidth]{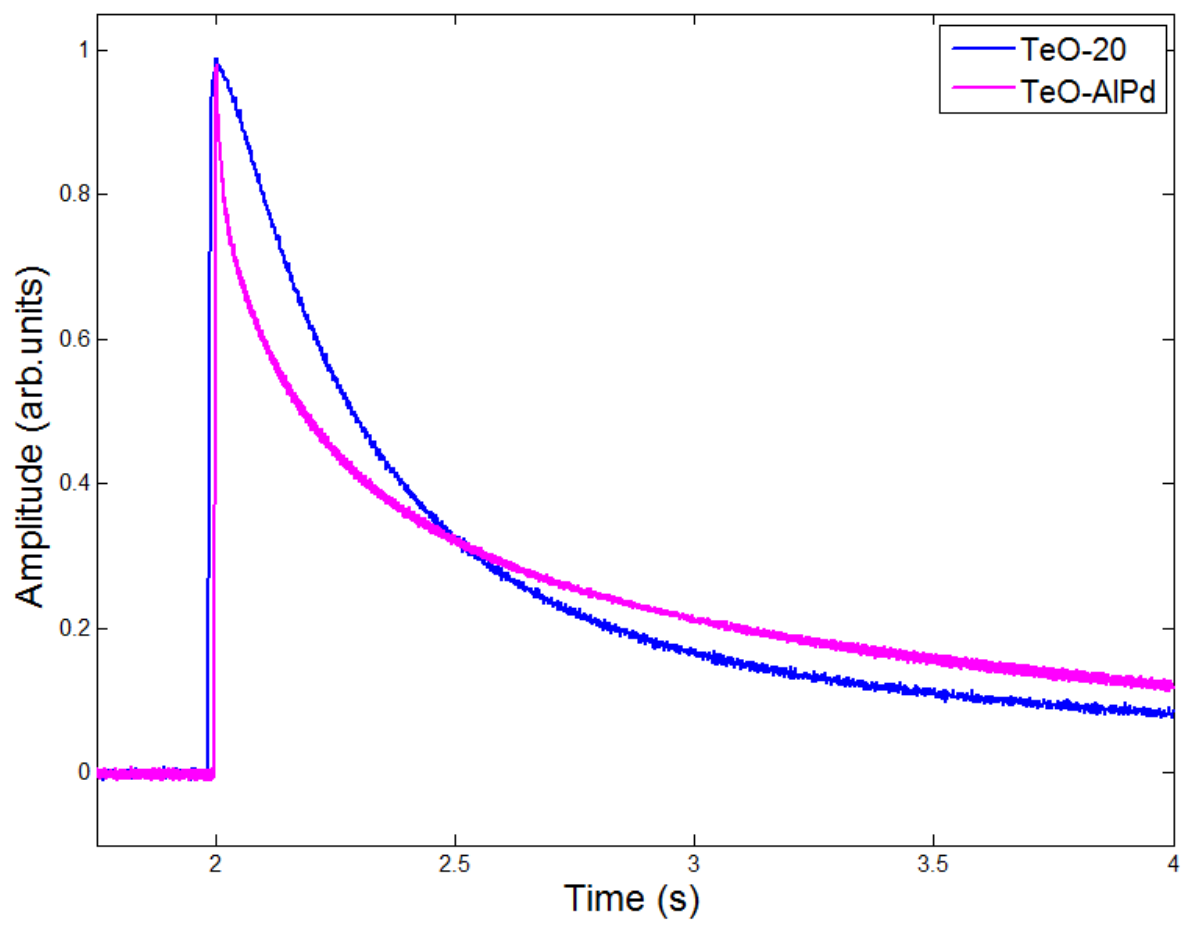}
\caption{Fragments of average signals of $^{130}$TeO$_2$ bolometers operated at a detector plate temperature of 17 mK (top left), 22 mK (top right), and 27 mK (bottom).}
\label{fig:TeO_meanpulse}
\end{figure}

After the characterization of the NTD-Ge sensors at the coldest detector plate temperature chosen (17 mK), we found that, for the optimal working point in terms of highest detector sensitivity, the measured resistances of the thermistors are too high --- $O$(100 M$\Omega$) --- for the optimal operation of the existing electronics. This is due to the significantly higher stress of the sensor coupled with UV-cured glue with respect to that observed for lithium molybdates \cite{CROSSdetectorStructure:2024}. Therefore, we characterized $^{130}$TeO$_2$ bolometers also at higher detector plate temperatures and polarized NTD-Ge with stronger currents reducing the resistances to a few-tens M$\Omega$. The detectors show a rather high sensitivity at 17 mK, $O$(100 nV/keV), which is reduced by a factor 10 at subsequent operation at warmer temperatures. It is worth emphasizing a comparatively small reduction (a factor of 2) in sensitivity of the metal-coated crystals (irrespective of single or 4-side coating with either a grid or a layer) with respect to the bare sample, in contrast to observations with Li$_2$MoO$_4$ crystals \cite{Khalife:2021}. 
A drastic decrease in the sensitivity at warmer working temperatures has a minor impact on the measured baseline noise after optimum filtering, which is around 2--5 keV RMS, similar to what was reported for Li$_2$MoO$_4$-based bolometers tested in this set-up in the same / similar detector structure \cite{CROSSdetectorStructure:2024}. It is worth noting the quite long thermal signals of the $^{130}$TeO$_2$ bolometers (figure~\ref{fig:TeO_meanpulse}) exhibiting a few seconds long descending part of the signals, similar to early reported response for such detector material, which are more than a factor 5 longer than the signals of the reference  Li$_2$$^{100}$MoO$_4$  bolometers \cite{CROSSdetectorStructure:2024}. At the same time, the internal $2\nu2\beta$-decay activity of $^{130}$Te in the enriched crystals is about two orders of magnitude lower than in the case of $^{100}$Mo (due to the difference in the half-lives, e.g. see recent results in \cite{Adams:2021a} and \cite{Augier:2023} respectively), thus it does not represent a notable background in form of pile-ups in contrast with the behaviour of $^{100}$Mo-enriched bolometers \cite{Chernyak:2012}. The performance of $^{130}$TeO$_2$ thermal detectors is summarized in table \ref{tab:TeO_enriched_performance}.

\begin{table}
\centering
\caption{Performance of $^{130}$Te-enriched TeO$_2$ bolometers (0.55 kg each) at different temperatures of the detector plate ($T_{plate}$). For each working point of the NTD thermistor (resistance $R_{NTD}$ at a given current $I_{NTD}$), we report the pulse-shape time constants ($\tau_{r}$ and $\tau_{d}$), the detector sensitivity ($A_s$; evaluated with 2615 keV $\gamma$ quanta), and the RMS baseline width (RMS$_{noise}$).}
\smallskip
\begin{tabular}{lccccccc}
\hline
Channel  & $T_{plate}$ & $R_{NTD}$ & $I_{NTD}$ & $\tau_{r}$ & $\tau_{d}$ & $A_s$ & RMS$_{noise}$ \\
~ & (mK) & (M$\Omega$)  & (nA) & (ms) &  (ms) & (nV/keV) & (keV)  \\
\hline
\hline
TeO-20      & 17  & 180 & 0.10 & 45  & 370  & 250    & 3.0       \\ 
~           & 22  & 26  & 0.55 & 7.0 & 360  &  38    & 2.7       \\ 
~           & 27  & 7.0 & 3.0  & 6.0 & 400  &   9    & 2.4       \\ 
\hline
TeO-Al      & 17  & 180 & 0.10 & 37  & 420  & 137    & 2.9       \\ 
~           & 22  &  25 & 0.55 & 8.0 & 420  &  18    & 3.6       \\ 
\hline
TeO-AlPd    & 17  & 170 & 0.10 & 36  & 600  & 101    & 4.7      \\ 
~           & 22  & 6.4 & 1.0  & 3.4 & 24   &  12    & 8.7      \\ 
~           & 27  & 3.0 & 3.0  & 3.0 & 550  &   7    & 4.6       \\ 
\hline
\end{tabular}
\label{tab:TeO_enriched_performance}
\end{table}

\begin{figure}
\centering
\includegraphics[width=0.7\textwidth]{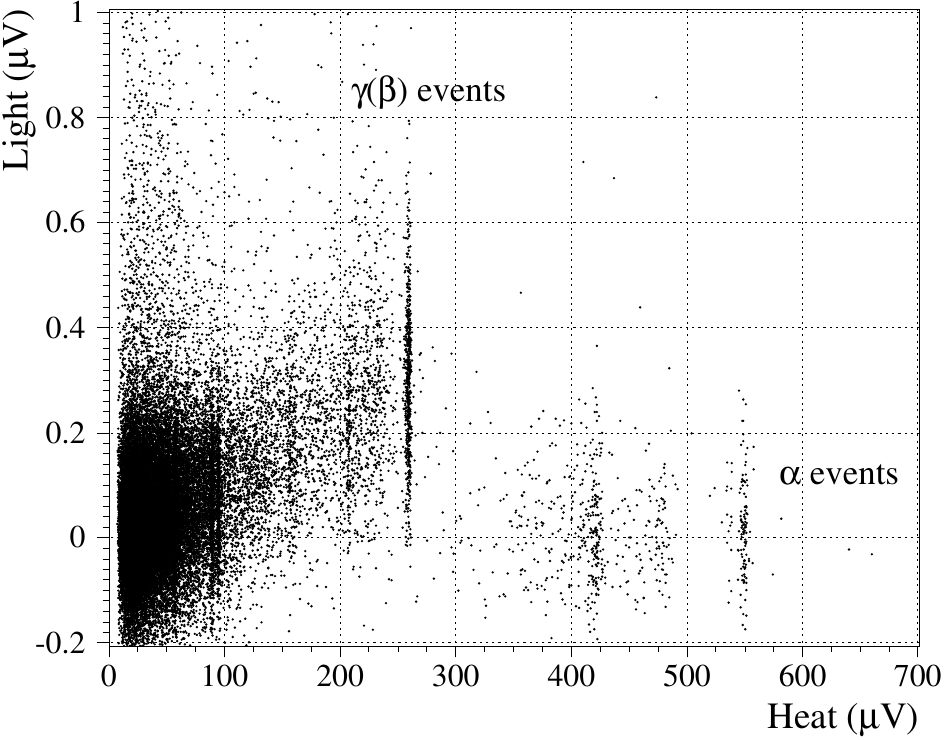}
\caption{Distribution of $^{130}$TeO$_2$-detected heat signals (TeO-AlPd) in coincidences with photon signals of the accompanied Ge bolometer (placed above the crystal) measured over 84-h $^{232}$Th calibration in the CROSS set-up at the LSC (Spain). The light detector was operational in the NTL amplification mode (50 V bias was applied on the Al electrode) allowing to reduce the baseline noise to ten(s) eV RMS. The population of $\gamma$($\beta$) events is mainly induced by $\gamma$s from the external $^{232}$Th source, while $\alpha$ particles detected are originated to either external U $\alpha$ source used (more broad distributions) or to internal contamination of the crystal (mainly by $^{210}$Po).}
\label{fig:TeO_enriched_Light_Heat}
\end{figure}

Detectors calibration and characterization is done with the help of a $^{232}$Th source, inserted inside the lead shielding. Coincidences between $^{130}$TeO$_2$ detectors and corresponding Ge bolometers have been established according to time properties (rise time) of their thermal signals. An example of a two-dimensional distribution of the amplitudes of the detected events is illustrated in figure \ref{fig:TeO_enriched_Light_Heat}. 
A band of events with a light signal amplitude clearly depending on the heat energy --- appreciable in figure \ref{fig:TeO_enriched_Light_Heat} --- corresponds to $\gamma$($\beta$) interactions in the $^{130}$TeO$_2$ absorber resulting to the emission of Cherenkov radiation. Clusters of high-energy particles with no detectable light signals correspond to events induced by $\alpha$ decays; a possible tiny scintillation of the TeO$_2$ material is not observed for $\alpha$-particles as the conditions of light collection are not favorable due to the absence of a reflective film around the crystal. This is also the main reason for only partial discrimination between $\alpha$ and $\gamma$($\beta$) populations\footnote{The impact of a single-side grid coating on the photon absorption and thus on photon output from the crystal is expected to be negligibly low, as can be concluded from studies of metal-coated lithium molybdate scintillating bolometers \cite{Bandac:2020, Bandac:2021}. On the contrary, the 4-side Al coating can lead to the efficient absorption of photons by the film resulting to a poor light output, as observed for lithium molybdate crystals \cite{CrossCupidTower:2023a}.}. We notice in fact that a much better discrimination was obtained with a device consisting of an even larger (50-mm cube) crystal placed inside a holder with a reflective foil \cite{Berge:2018}, characterized by a similar-performance but smaller light detector. The discrimination will be improved by using an optimized geometry of the Al electrode with a better Ge-wafer surface coverage, which will allow us to improve the amplification by a factor 2 \cite{CUPIDalternativeStructure:2024}. Moreover, a further enhancement in the NTL gain can be achieved by facing grids to the crystal surface and by applying a twice higher electrode bias \cite{CUPIDalternativeStructure:2024}.

\begin{figure}
\centering
\includegraphics[width=0.8\textwidth]{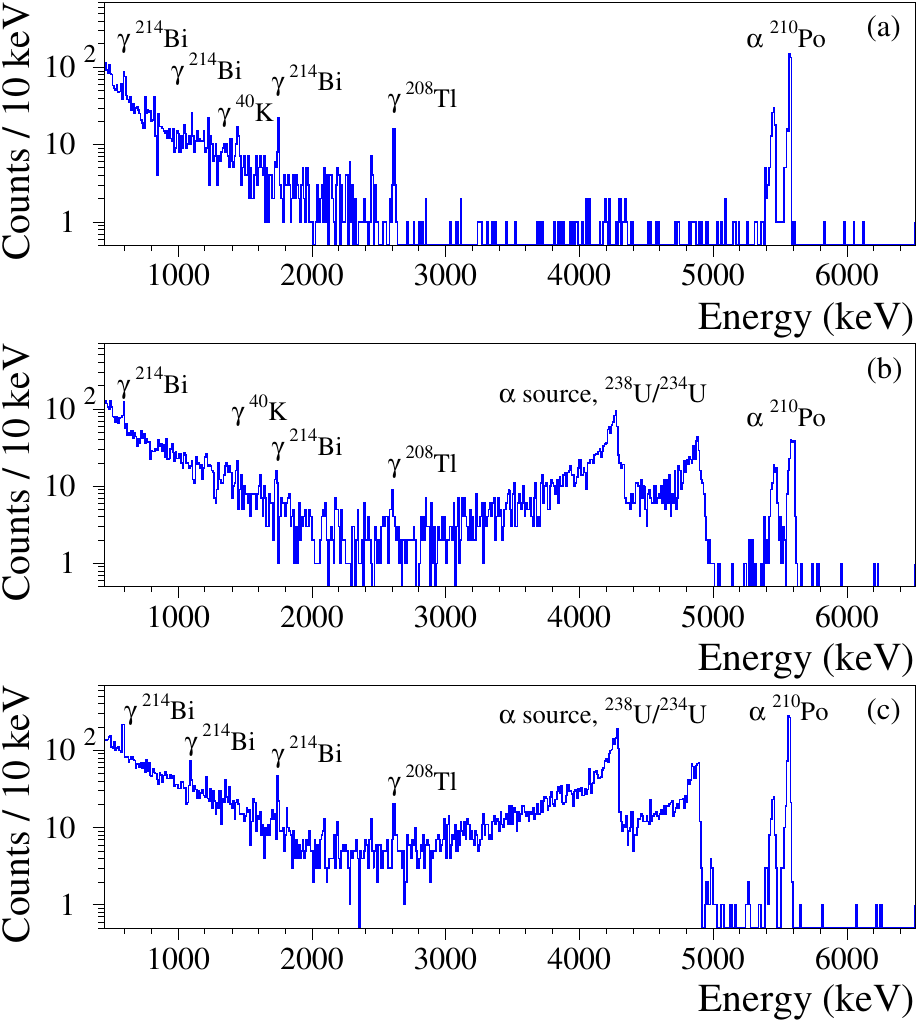}
\caption{Energy spectra of events detected by the 0.55-kg $^{130}$TeO$_2$ bolometers TeO-20 (a), TeO-AlPd (b), and TeO-Al (c). Both metal-coated bolometers were additionally irradiated by a close $\alpha$ source of $^{238}$U/$^{234}$U. The data were acquired in the CROSS set-up at the Canfranc underground laboratory in the following conditions: (a) and (b) correspond to measurements at 27 mK over 116 h (reference date is Nov. 2023); (c) represents a 168-h-long dataset collected at 17 mK (reference date is Mar. 2023). The origin of the most intense $\gamma$ and $\alpha$ peaks observed in the spectra is labeled.}
\label{fig:TeO_enriched_Bkg_spectra}
\end{figure}

The $^{130}$TeO$_2$ bolometers were operated at a higher temperature --- regulated at 27 mK --- where the performance is expected to be better considering the high NTD resistances. In these conditions, we performed a relatively long measurement (116 h) without the external $^{232}$Th $\gamma$ source, attempting to probe the radiopurity level of the samples. In the case of the TeO-Al sample, we used such type of ``background'' data collected at 17 mK with a similar duration (168 h), realized 7.5 months before the measurement at the warmest working point. 
The energy spectra of the three investigated detectors showing the region of interest for $\gamma$, $\beta$, and $\alpha$ spectroscopy are depicted in figure \ref{fig:TeO_enriched_Bkg_spectra}. The experiment was not supplied by the air flow with low radon content, thus the presented data exhibit a significant $\gamma$ background induced by radioactive daughters (mainly $^{214}$Bi) of radon decay, as demonstrated in \cite{CROSSdeplLMO:2023}. In the $\alpha$ region, we observed only a single peak-like structure corresponding to the internal $^{210}$Po contamination, while a nearby peak, shifted by 100 keV to lower energies is traced to surface pollution of detector materials by $^{210}$Po ($^{210}$Pb). In addition, as already mentioned above, data of TeO-AlPd and TeO-Al detectors exhibit two broad peaks from an external $\alpha$ source used to irradiate the crystal. It is interesting to note that the energy of the ``bulk'' $^{210}$Po peak, calibrated by 2614.5 keV $\gamma$ quanta, is found about 3\% higher than the nominal value (5407.5 keV) for all data and crystals, except 17 mK measurements with the TeO-20 detector showing a $\sim$1\% deficit in the estimated energy. The observed miscalibration of the $^{210}$Po $\alpha$ peak is similar to early reported data of TeO$_2$ bolometers (e.g. a quenching factor of $\sim$1.02 \cite{Alessandrello:1997,Berge:2018} and $\sim$0.99 \cite{Bellini:2010}) and it can be explained by non-linearity of the bolometric response, as demonstrated in \cite{Bellini:2010}, where the quenching factor 1.0076(5) was measured for the 2.3 MeV $\alpha$ peak of $^{147}$Sm. This comparison additionally shows a strong impact of measurements and data analysis on the estimate of quenching factors for alpha particles and nuclear recoils, similarly to such studies with other detector materials \cite{Tretyak:2010,Tretyak:2014}. 

The activity of $^{210}$Po detected in the $^{130}$Te-enriched crystals, as shown in figure \ref{fig:TeO_enriched_Bkg_spectra}, is 1.6(1), 0.8(1), and 2.5(1) mBq/kg for the sample TeO-20, TeO-AlPd, and TeO-Al respectively. This is a conservative estimate of the $^{210}$Po bulk contamination, because it includes a contribution of the crystal surface radioactivity, while a comprehensive background model is needed to scrutinize the bulk and surface components, e.g. as done in \cite{Alduino:2017a}. For the same reference date (Nov. 2023) of the TeO-20 and TeO-AlPd samples, corresponding to the end of the experiment described here, the $^{210}$Po activity in the TeO-Al crystal is expected at 0.8 mBq/kg. At the beginning of the experiment (Mar. 2023), the $^{210}$Po activity in TeO-20 and TeO-AlPd, estimated from the same 168-h-long dataset, is 4.5(1) and 2.4(1) mBq/kg respectively, in agreement with the expectations based on 27 mK data (Nov. 2023). Thus, the $^{210}$Po content is a few times lower than that in the precursor samples with natural Te isotopic abundance. This confirms the efficiency of the additional purification applied in the process of the $^{130}$TeO$_2$ crystal development. Taking into account that the $^{228}$Th (5520 keV) and $^{220}$Rn (5590 keV) peaks are below the detection limit, we can conclude that the activity of $^{228}$Th and $^{226}$Ra in the crystal bulk is expected to be at least two orders of magnitude lower than the observed $^{210}$Po content. Altogether, this proves a high internal radiopurity of the $^{130}$TeO$_2$ crystals developed for the CROSS experiment, which can also comply demands of next-generation experiments.

\section{Conclusions}

In the framework of the CROSS project, we developed a production chain of tellurium dioxide crystals from highly purified $^{130}$Te-enriched powder (with 93\% of $^{130}$Te in the isotopic composition of Te). The directional solidification method has been used for the additional purification of the powder, reducing the concentration of impurities by a factor 10, down to a few ppm level in total. The purest part of the ingot is determined by the segregation profiles of impurities (surface / bulk of the ingot); the yield is around 80\%. 
The capability of the selected crystal-growth company (using Czochralski method) to supply high quality TeO$_2$ samples with the required internal radiopurity (e.g. below 10 mBq/kg activity of $^{210}$Po) has been validated by the bolometric results obtained above ground with two small-size samples (20 $\times$ 20 $\times$ 10 mm), produced from a powder with natural Te isotopic abundance and no additional purification. After that, we proceeded with the fabrication of six large $^{130}$TeO$_2$ crystals with the size of 45 $\times$ 45 $\times$ 45 mm and a mass of 0.55 kg each. The isotopic abundance of $^{130}$Te remains very high in the crystals ($\sim$91\%) thanks to the adopted Czochralski technique (with a small seed), avoiding a significant dilution of the $^{130}$Te-enriched material in the growth seed.

Four over six $^{130}$TeO$_2$ samples have been operated as thermal detectors in the CROSS low-background set-up at the Canfranc underground laboratory (Spain), using a newly developed detector structure for the CROSS experiment. Half of the samples tested were coated with either Al film (on 4 sides) or with AlPd grid (on 1 side). The detectors show high bolometric and spectrometric performance, as well as a low internal contamination by $\alpha$-active radionuclides from U/Th chains. The only detected contamination is ascribed to $^{210}$Po with an activity of around 1 mBq/kg (reference date is Nov. 2023), which is a few times lower than in precursors with natural Te composition and no additional purification. Activities of the most harmful radionuclides for double-beta decay search experiments, $^{228}$Th and $^{226}$Ra from $^{232}$Th/$^{238}$U families, are expected to be at least two orders of magnitude lower than that of $^{210}$Po (i.e. below $\sim$10 $\mu$Bq/kg). A high radiopurity level of newly developed large-volume $^{130}$TeO$_2$ crystals fully satisfies the CROSS demands and, potentially, requirements of next-generation experiments.

\acknowledgments

This work is supported by the European Commission (Project CROSS, Grant No. ERC-2016-ADG, ID 742345), by the Agence Nationale de la Recherche (ANR France; Project CUPID-1, ANR-21-CE31-0014), and US National Science Foundation (NS~1614611). We acknowledge also the support of the P2IO LabEx (ANR-10-LABX0038) in the framework ``Investissements d'Avenir'' (ANR-11-IDEX-0003-01 -- Project ``BSM-nu'') managed by ANR, France. 
This work was also supported by the National Research Foundation of Ukraine under Grant No. 2023.03/0213 and by the National Academy of Sciences of Ukraine in the framework of the project ``Development of bolometric experiments for the search for double beta decay'', the grant number 0121U111684. 
Russian and Ukrainian scientists have given and give crucial contributions to CROSS. For this reason, the CROSS collaboration is particularly sensitive to the current situation in Ukraine. The position of the collaboration leadership on this matter, approved by majority, is expressed at \href{https://a2c.ijclab.in2p3.fr/en/a2c-home-en/assd-home-en/assd-cross/}{https://a2c.ijclab.in2p3.fr/en/a2c-home-en/assd-home-en/assd-cross/}.


\end{document}